\documentclass[a4paper,12pt]{article}

\usepackage[mathscr]{eucal}
\usepackage{amsmath,amsfonts,amssymb,amsthm}
\usepackage{times}
\usepackage{hyperref}

\advance\textheight\topskip
\numberwithin{equation}{section} \makeatletter
\@addtoreset{equation}{section}

\newcommand{\ul}[1]{{\underline{#1}}}

\renewcommand{\tilde}{\widetilde}
\renewcommand{\hat}{\widehat}

\newcommand{\bref}[1]{\textbf{\ref{#1}}}

\newcommand{\gh}[1]{\mathrm{gh}(#1)}

\newcommand{\dd}{\partial}
\renewcommand{\d}{\partial}

\newcommand{\tensor}{\otimes}

\renewcommand{\geq}{\,{\geqslant}\,}
\renewcommand{\leq}{\,{\leqslant}\,}

\newcommand{\binner}[2]{%
  {\langle}\kern-4.15pt{\langle}#1{,}\,#2{\rangle}\kern-4.15pt{\rangle}}
\newcommand{\commut}[2]{[#1{,}\,#2]}

\newcommand{\half}{\mathchoice{%
    \ffrac{1}{2}}{\frac{1}{2}}{\frac{1}{2}}{\frac{1}{2}}}

\newcommand{\ffrac}[2]{\raisebox{.5pt}%
  {\footnotesize$\displaystyle\frac{#1}{#2}$}\kern1pt}

\newcommand{\derham}{\boldsymbol{d}}

\newcommand{\dl}[1]{\mathchoice{\ffrac{\dd}{\dd #1}}{\frac{\dd}{\dd
      #1}}{\ffrac{\dd}{\dd #1}}{\ffrac{\dd}{\dd #1}}}

\newcommand{\manifold}[1]{\mathscr{#1}}
\newcommand{\manX}{\manifold{X}}

\newcommand{\Liealg}{\mathfrak} 
\newcommand{\algg}{\Liealg{g}}

\newcommand{\fR}{\mathbb{R}}

\newcommand{\NN}{\mathbb{N}}

\def\cH{\mathcal{H}}

\def\sd{S^\dagger}
\def\bsd{\bar S^\dagger}

\def\BGST{Barnich:2004cr}
\def\BGadS{Barnich:2006pc}

\def\BG-Poincare{Barnich:2009jy}
\def\Fedosov-book{Fedosov:1996fu}

\usepackage{youngtab}

\newcommand{\barBox}{\overset{-}{\Box}{}}
\newcommand{\tildeBox}{\overset{\sim}{\Box}}
\newcommand{\diag}{\operatorname{diag}}

\begin{document}

%\begin{titlepage}
\begin{center}
\textbf{\Large{Boundary values of mixed-symmetry massless fields in AdS space
}}

% \textbf{\Large
% Boundary values of massless AdS fields of arbitrary symmetry type
% }}

\vspace{1.5cm}

%\vspace{1.4cm}

  {\large {Alexander Chekmenev ~~and~~~ Maxim Grigoriev}}

\vspace{.5cm}

  %
%
%\end{titlepage}
%\maketitle

%\pagebreak

%\address

\textit{Tamm Theory Department,
 Lebedev Physics Institute\\
 Leninsky prospect 53, 119991 Moscow, Russia}

 \vspace{0.5cm}

\textit{
Moscow Institute of Physics and Technology, Dolgoprudny,\\
141700 Moscow region, Russia
}

\end{center}

\vspace{1cm}

\begin{abstract}
We elaborate on the ambient space approach to boundary values of $AdS_{d+1}$ gauge fields and apply it to massless fields of mixed-symmetry type. In the most interesting case of odd-dimensional bulk the respective leading boundary values are conformal gauge fields subject to the invariant equations. Our approach gives a manifestly conformal and gauge covariant formulation for these fields. Although such formulation employs numerous auxiliary fields, it comes with a systematic procedure for their elimination that results in a more concise formulation involving only a reasonable set of auxiliaries, which eventually (at least in principle) can be reduced to the minimal formulation in terms of the irreducible Lorentz tensors. The simplest mixed-symmetry field, namely, the rank-3 tensor associated to the two-row Young diagram, is considered in some details.

\end{abstract}
\thispagestyle{empty}

\vspace{1cm}

\setcounter{tocdepth}{2}
\tableofcontents

% % % % % % % % % % % % % % % % % % % % % % % % % % % % % % % % % % % % % % % %
%                                 INTRODUCTION
%

\section{Introduction}
At the kinematical level the celebrated AdS/CFT duality (for a review see~\cite{Aharony:1999ti}) is heavily based on the notion of boundary values. For an AdS field there are typically two options to prescribe asymptotic behavior in a way compatible with AdS isometries (i.e. $o(d,2)$ invariance). These correspond to leading and sub-leading boundary values. While sub-leading  boundary values are identified with the conformal operators of the boundary theory, the leading ones are associated with the respective sources (see e.g.~\cite{Skenderis:2002wp}). 

In the case where the bulk AdS space is odd-dimensional, the boundary values of (partially)-massless fields can be subject to invariant conformal equations. The respective action typically shows up as a logarithmically-divergent part in the effective action. In the case of general unitary totally-symmetric fields this was demonstrated in~\cite{Metsaev:2009ym,Metsaev:2013wza} in a gauge covariant way. Similar analysis has been performed for a particular ``hook-type'' mixed-symmetry field~\cite{Alkalaev:2012ic} (see also~\cite{Alkalaev:2012rg}). As far as general mixed-symmetry fields are concerned, only the light-cone  approach~\cite{Metsaev:2015rda} is available in the literature so far. 

An alternative point of view on the equations of motion satisfied by the leading boundary value is to treat them as an obstruction to extending the off-shell boundary value to a bulk on-shell field configuration. This point of view was recently put forward in~\cite{Bekaert:2012vt,Bekaert:2013zya}, resulting in a general method to study boundary values in a manifestly $o(d,2)$-invariant and gauge covariant way. This is achieved by using the ambient space construction along with the BRST and jet-space techniques. More precisely,
we employ the parent formulation approach~\cite{\BGST,Barnich:2006pc,Barnich:2010sw,Grigoriev:2010ic} which incorporates both these techniques. In contrast to the usual approach the conservation condition for the subleading boundary value as well as the conformal equations of motion for the leading one arise in exactly the same way. Using this method the gauge covariant analysis of the boundary values of totally symmetric (partially-)massless fields and associated conformal equations has been performed in~\cite{Bekaert:2013zya}.

In this work the boundary values of bosonic gauge fields in $AdS_{d+1}$ space of arbitrary symmetry type are studied at the level of equations of motion. It turns out that the method of \cite{Bekaert:2012vt,Bekaert:2013zya} extends smoothly to this case. More precisely, we limit ourselves to unitary massless fields originally studied in~\cite{Metsaev:1995re,Brink:2000ag,Alkalaev:2003qv} (see also~\cite{Skvortsov:2009zu,Alkalaev:2009vm,Skvortsov:2009nv,Boulanger:2008up,Zinoviev:2009gh,Campoleoni:2012th}).

We are mainly focused on the case of the even-dimensional boundary where our approach produces the manifestly conformal formulations for a rather general class of bosonic mixed-symmetry gauge fields. Such formulations for mixed symmetry gauge fields were not known in the literature to the best of our knowledge. As a price for the manifest conformal invariance, the formulation we arrived at involve a plenty of auxiliary fields. However, just like in the totally symmetric case considered in~\cite{Bekaert:2012vt,Bekaert:2013zya}, a systematic and explicit procedure to eliminate auxiliaries and arrive at the formulation in terms of Lorentz irreducible fields is available. Remarkably, keeping some of the auxiliary fields results in the formulation closely related to that proposed by Metsaev~\cite{Metsaev:2007rw}. Mention that Lagrangians of generic (including mixed-symmetry) conformal gauge fields were originally obtained by Vasiliev~\cite{Vasiliev:2009ck} using a different framework.

\section{Ambient space approach to boundary values}
\label{sec:bound-gen}
\subsection{(Critical) AdS scalar and its boundary values}
\label{sec:scalar}

To illustrate the ambient space approach to boundary values let us review in some details the simplest case of a scalar field. For more details see~\cite{Bekaert:2012vt,Bekaert:2013zya} and references therein. A scalar field $\varphi$ of mass $m$ on $AdS_{d+1}$ is described by the following equation of motion:
\begin{equation}
\label{AdSkg}
 (\nabla^2-m^2)\varphi=0\,.
\end{equation} 

It is well known that AdS spacetime can be seen as a hyperboloid embedded in the ambient space which is a flat pseudo-Euclidean space $\fR^{d,2}$. Let $X^A, A = 0,\ldots,d+1$ be Cartesian coordinates on $\mathbb R^{d,2}$ and $\eta_{AB} = \diag\{-+\dots+-\}$ the metric, then the embedding reads explicitly as
\begin{equation}
\eta_{AB} X^A X^B = -1\,.
\end{equation}
Note that AdS isometries lift to ambient pseudo-orthogonal transformations and hence
are linear transformations in terms of ambient space coordinates. That is one of the main reasons why ambient approach is useful in describing AdS fields.

In terms of the ambient space the above scalar equation of motion \eqref{AdSkg} can be represented as
\begin{equation}
\label{ads-scalar}
 (\d_X\cdot \d_X)\, \Phi=0\,, \qquad (X\cdot \d_X+\Delta)\Phi=0\,,
\end{equation} 
where $m^2=\Delta(\Delta-d)$ and $\cdot$ denotes the invariant contraction of the ambient indices e.g. $X\cdot \d_X=X^A\dl{X^A}$. 
The field $\Phi$ is a lift of $\varphi$ defined on the hyperbolid to the ambient space, i.e. $\Phi|_{X^2=-1}=\varphi$. It is defined in the vicinity of the hyperboloid $X^2=-1$ which thanks to the second equation is the same as defining $\Phi$ in the domain $X^2<0$. Note that there are in general two possible values of $\Delta$ associated with the same mass: $\Delta_\pm$, $\Delta_-\leq \Delta_+$.

The conformal boundary of $AdS_{d+1}$ can be identified with the projectivization of the hypercone $X^2=0$ or, more precisely, the manifold of null rays. One typically identifies this manifold with a submanifold $\manX$ of the hypercone such that each ray corresponds
to a point of $\manX$ and vice versa (here we ignore global geometry subtleties).  Manifold $\manX$ is not equipped with Riemannian metric but rather with a conformal structure. Identifying the manifold of null rays with $\manX$ gives a Riemannian metric -- the pullback of the ambient metric to $\manX$. However, a different identification leads to a conformally equivalent metric (related by the Weyl transformation $g_{\mu\nu}(x)\to \Lambda^2(x)g_{\mu\nu}(x)$). A useful choice for $\manX$ is the surface $X^+=1$, $X^2=0$. With this choice the section is identified with the $d$-dimensional Minkowski space.

The boundary value of $\Phi$ is the value of $\Phi$ on the hypercone $X^2=0$, which thanks to the second equation in~\eqref{ads-scalar} is uniquely determined by the value of $\Phi$ on $\manX$ and can be seen as defined on the manifold of null rays. Note that $\Phi$ is defined on $X^2<0$ only so that for a generic solution the boundary value may be ill-defined. 
In this way boundary values depend on the choice of $\manX$ and in more geometrical language are densities on $\manX$ rather than scalar functions. For simplicity, in what follows we fix $\manX$ to be $X^+=1$, $X^2=0$ and hence treat boundary values as functions on $\manX$.

More precisely, suppose we are given with  a solution $\varphi$ to \eqref{AdSkg}. Its leading (respectively sub-leading) boundary value is defined as follows: first one lifts $\varphi$ to a solution $\Phi$ to~\eqref{ads-scalar} with $\Delta=\Delta_-$ (respectively $\Delta=\Delta_+$). This lift is unique thanks to the second equation in~\eqref{ads-scalar}. The boundary value is then defined as $\phi(X)=\lim_{X\to \manX}\Phi(X)$ or, more precisely, with our choice of $\manX$ and using ambient coordinates $X^\pm,X^a$ one has
\begin{equation}
 \phi(X^a)=\lim_{X^-\to -\half X^aX_a} \Phi(X^+=1,X^-,X^a) \,.
\end{equation} 

If well-defined, (sub)leading boundary values can be considered as certain conformal fields. In the case of the scalar field the subleading boundary value is always an off-shell conformal scalar field, i.e. the scalar of conformal weight $\Delta_+$ not subject to any equations. Configurations of such scalar are known to form an irreducible module of $o(d,2)$. For generic $m^2$ the same happens for the leading boundary value which has conformal weight $\Delta_-$. Again, this can be seen as the off-shell conformal scalar of weight $\Delta_-$.

For $\Delta$ generic or $\Delta=\Delta_+$ boundary values are described by the same ambient space equations~\eqref{ads-scalar} but for $\Phi$ defined in the vicinity of the hypercone~\cite{Bekaert:2012vt,Bekaert:2013zya}. An important subtlety occurs if the AdS scalar is \textit{critical}. This is the case where $\Delta_-=\frac{d}{2}-\ell$ with $\ell$ positive integer. In this case the leading boundary value is not off-shell and the space of solutions to~\eqref{ads-scalar} in the vicinity of $X^2=0$ is not irreducible. More precisely, for $\Delta_-=\frac{d}{2}-\ell$ the space contains an invariant subspace of solutions of the form $(X^2)^\ell\alpha(X)$, where $\alpha$ satisfies~\eqref{ads-scalar} with $\Delta=\Delta_+=\frac{d}{2}+\ell$, i.e. $\alpha$ corresponds to the subleading boundary value. This can be interpreted as a gauge equivalence so that the space of inequivalent solutions coincides with the space of leading boundary values which in the case at hand are subject to polywave equation $\Box_0^\ell \phi=0$. Here and below $\Box_0=\dl{x^a}\dl{x_a}$.

As a byproduct of the above construction one gets a manifestly conformal description of the conformal equations $\Box_0^\ell \phi=0$. Indeed, supplementing the ambient system~\eqref{ads-scalar} with the above gauge equivalence gives a manifestly $o(d,2)$-invariant ambient system
\begin{equation}
\label{conf-scalar}
 (\d_X\cdot \d_X) \Phi=0\,, \qquad (X\cdot \d_X+\frac{d}{2}-\ell)\Phi=0\,, \qquad \Phi \sim \Phi+(X\cdot X)^{\ell}\alpha\,,
\end{equation} 
which in the vicinity of $X^2=0$ is equivalent to the polywave equation. This system is generalized~\cite{Bekaert:2012vt,Bekaert:2013zya} to the case of totally-symmetric 
(partially-)massless fields on $AdS_{d+1}$, giving a manifestly conformal description of their boundary values ((generalized) Fradkin--Tseytlin conformal gauge fields). As we are going to see shortly it can be also generalized to the case of mixed-symmetry massless fields and their associated conformal fields on $\manX$.

\subsection{Weyl module(s) for the ambient system}

Given linear equations of motion an important object is a vector space $\cH_0$ of its (gauge inequivalent) solutions in the space of formal power series around a fixed space time point. This is just a stationary surface of the equations of motion at this spacetime point, seen as a linear space rather than a submanifold of the respective jet space. Let us recall that a stationary surface is a submanifold of the jet-space singled out by the prolonged equations. It is this manifold that underlies the invariant definition of the differential equation~\cite{Vinogradov:1978} (for a modern review see e.g.~\cite{Krasil'shchik:2010ij}).

In the case where the system is invariant under one or another spacetime symmetry group $G$ that acts  transitively on the spacetime, $\cH_0$ is clearly a module over the group and modules associated to different spacetime points are isomorphic. This module is well known in the unfolded approach~\cite{Vasiliev:1980as,Lopatin:1987hz,Vasiliev:2005zu} as \textit{Weyl} module. In this case this module contains all information of the starting point equations in the sense that the system can be completely reconstructed in terms of $\cH_0$-valued fields and the flat $\algg$-covariant derivative (which is naturally defined on the spacetime because it can be seen as a $G$-coset). 

More precisely, if $\nabla$ is a natural flat $\algg=Lie(G)$-connection on $G$-coset and $\psi$ an $\cH_0$-valued field it turns out that the following system of equations
\begin{equation}
\nabla \psi =0\,, 
\end{equation}
is equivalent to the starting point linear equations of motion through the elimination of auxiliary fields. Such formulation is known as the unfolded form of the system.

For the ambient system~\eqref{ads-scalar} the Weyl module is not unique because the symmetry group $O(d,2)$ doesn't act transitively on $\fR^{d,2}$. In particular the Weyl module at $X^2=0$  (i.e. on the hypercone) in general differs from that at $X^2=-1$ (i.e. on the hyperboloid) because these are different orbits of $O(d,2)$. To anticipate, the difference is precisely that between the bulk field and its boundary value seen as a conformal field. Note, however, that modules at $X^2=-R^2$ are isomorphic for all $R>0$. It is instructive to  compare the modules at $X^2=-1$ and $X^2=0$ explicitly. 

\subsubsection{Weyl module at $X^2=-1$}
\label{sec:weyl-ads}

For  $X^2=-1$ let us pick a point with coordinates $X^A=V^A\equiv\delta^A_{d+1}$. Formal series around $X^A$ are written as $\Phi(V+Y)$ (as $V$ is fixed we simply write $\Phi(Y)$) and the equations~\eqref{ads-scalar} take the form
\begin{equation}
\label{ads-conf-scalar}
 \d_Y\cdot \d_Y \Phi=0\,, \qquad \left((V+Y)\cdot \d_Y+\Delta\right)\Phi=0\,, 
\end{equation} 
Using notation $y^n=Y^n$, $n=0,\ldots ,d$ and $z=Y^{d+1}$ the second equation uniquely determines $z$-dependence of $\Phi$ in terms of its $z$-independent component. Taking the second equation into account shows that there is 1:1 correspondence between solution to~\eqref{ads-conf-scalar} and the $z$-independent elements annihilated by
$\dl{y^n}\dl{y_n}$ (i.e. harmonic elements).

In more details, given $\phi(y)$ such that $\dl{y^n}\dl{y_n}\phi=0$ equations~\eqref{ads-conf-scalar} can be solved order by order in $z$ as
\begin{multline}
 \Phi(y,z)=\phi(y)~-z(n+\Delta)\phi(y)~+\half z^2 (n+\Delta+1)(n+\Delta)\phi(y)~+
 \\
 ~+y^2\frac{(n+\Delta+1)(n+\Delta)}{2(d+1+2n)}\phi(y)~+\ldots
\end{multline}
where dots denote terms of either at least cubic order in $z$ or proportional to $(y^2)^2$ and $n=y^n\dl{y^n}$. The solution is unique for a given harmonic (i.e. with traceless Taylor coefficents) $\phi(y)$. More precisely, given a harmonic $\phi(y)$ there exist a unique solution to~\eqref{ads-conf-scalar}
satisfying $\Pi(\Phi|_{z=0})=\phi$, where $\Pi$ denotes the projector to the traceless component. 

In terms of $\Phi(Y)$ the $o(d,2)$ generators~\eqref{od2-gen} are given by:
\begin{equation}
\label{od2-gen}
J_{AB}\Phi=\left((V_A+Y_A)\dl{Y^B}-(V_B+Y_B)\dl{Y^A}\right)\Phi \,.                                                            \end{equation} 
Because solution to \eqref{ads-conf-scalar} are in 1:1 correspondence with harmonic $\phi(y)$ it is easy to write the action of $J_{AB}$ in these terms. Namely
\begin{equation}
 J_{AB}\phi=\Pi\Big((J_{AB}\Phi)|_{z=0}\Big)\,,
\end{equation} 
where $\Phi$ denotes the unique solution to \eqref{ads-conf-scalar} such that $\Pi(\Phi|_{z=0})=\phi$.

In particular for $\hat P_n=J_{nz}=y_n\dl{z}+(z+1)\dl{y^n}$ one finds
\begin{multline}
\label{ads-trans}
\hat P_n \phi=\Pi((P_n\Phi)|_{z=0})=\Pi\left(
-y_n(n+\Delta)\phi
+\dl{y^n}\phi
+y^n\frac{(n+\Delta+1)(n+\Delta)}{d+1+2n}\phi
\right)=
\\
=\dl{y^n}\phi-\Pi\left(y^n\frac{(n+\Delta)(n+d-\Delta)}{d+1+2n}\phi\right)
\end{multline}
This defines the action of AdS translation on the Weyl module. As for Lorentz rotations these are simply represented by
\begin{equation}
 J_{nm}\phi=(y_n\dl{y^m}-y_m\dl{y^n})\phi\,.
\end{equation} 

For $\Delta$ generic the module is clearly irreducible. For $\Delta =-N$ and  $\Delta = d+N$ where $N\in \NN_0$ (recall that $\NN_0$ denotes nonnegative integers) the module contains a finite-dimensional submodule of elements of homogeneity not exceeding $N$. Factoring out the submodule one arrives at the irreducible module.

It is worth mentioning that the form of the coefficient in~\eqref{ads-trans} tells us that the structure of the module is invariant under $\Delta \to d-\Delta$. This confirms that the choice between $\Delta_+$ and $\Delta_-$ is irrelevant if we are only concerned with AdS fields.

To complete the discussion of AdS Weyl module let us mention that the module was originally arrived at from different perspective in~\cite{Shaynkman:2000ts} (analogous modules were already in~\cite{Vasiliev:1980as,Lopatin:1987hz}). The derivation we have just given is the straightforward generalization of that from~\cite{Barnich:2006pc} (see also~\cite{Alkalaev:2009vm,Alkalaev:2011zv}) to which we refer for further technical details and generalizations. 

\subsubsection{Weyl module at $X^2=0$}
\label{sec:Weyl-conf}
Now we study the Weyl module for  the boundary value of the scalar of mass $m$ and boundary behavior determined by $\Delta_+$ or $\Delta_-$. In contrast to AdS module the choice between $\Delta_+$ and $\Delta_-$ is important.

To describe the Weyl module we pick a point of the hypercone $X^2=0$ with coordinates 
$X^A=V^A$ and solve~\eqref{ads-conf-scalar}. The convenient choice is $V^+ = 1, V^- = V^a = 0$, $a = 0, \dots, d-1$ where we make use of the conventional light-cone basis $E_+,E_-,E_a$ such that the nonvanishing components of $\eta$ are $\eta(E_+,E_-)=1, \eta(E_a,E_b)=\eta_{ab}$.

Let us denote light-cone coordinates as
\begin{equation}
	Y^+ = v,\quad
	Y^- = u,\quad
	Y^a = y^a
\end{equation}
and write $\Phi$ as $\Phi(u,v,y) = \sum_{i,j} \frac{u^i}{i!} \frac{v^j}{j!} \psi^i_j(y)$. In terms of components equations \eqref{ads-conf-scalar} take the following form
\begin{equation}
\label{iter}
	(n + \Delta + i + j) \psi^i_j + \psi^i_{j+1} = 0,\quad
	2 \psi^{i+1}_{j+1} + \partial^a\partial_a \psi^i_j = 0\quad
	i,j \ge 0.
\end{equation}
It follows that unless $-\Delta\notin \mathbb N_0$ all the components can be uniquely expressed in terms of $\psi_0$. We limit ourselves to this case. For $-\Delta\in \mathbb N_0$ the Weyl module in addition contains a finite-dimensional submodule which in the ambient space terms is given by harmonic homogeneous polynomials in ${Y^\prime}^A\equiv Y^A+V^A$ of degree $-\Delta$. The group-theoretical modules associated to the conformal scalar of weight $-\Delta\in \mathbb N_0$ were studied in~\cite{Basile:2014wua}.

\subsubsection{$\Delta \ne 0, -1, -2, \ldots$}
In this case the module is parameterized by $\phi=\psi^0_0(y)$. Using~\eqref{iter} it's easy to write $o(d,2)$ generators \eqref{od2-gen} in terms of $\phi$
\begin{align}
  J_{ab} &\phi =
  \left(
    y_a \frac{\partial}{\partial y^b}
  - y_b \frac{\partial}{\partial y^a}
  \right) \phi,\\
  J_{+-} &\phi =
  \left(
    n + \Delta
  \right) \phi,\\
  J_{+a} &\phi =
  y_a
  \left(
    n + \Delta
  \right) \phi,\\
  J_{-a} &\phi =
  \left(
    \frac{\partial}{\partial y^a}
  - y_a \frac{1}{2(n + \Delta + 1)} \frac{\partial}{\partial y_c} \frac{\partial}{\partial y^c}
  \right) \phi,
\end{align}
where $n=y^a\dl{y^a}$.

It follows from the above explicit form that for $-\Delta \notin \mathbb N_0$ we are dealing with a generalized Verma module. This can also be seen as follows: for $-\Delta \notin \mathbb N_0$ generators  $J_{+a}$ act freely and, moreover, constants form a lowest weight subspace so that the module is freely generated by $J_{+a}$ from the lowest weight subspace.

Let us recall (for more details see e.g.~\cite{Dixmier}) that the structure of a generalized Verma module is encoded in its singular vectors, i.e. in our case eigenvectors of $J_{+-}$ (more generally, subspaces)  annihilated by $J_{-a}$. A singular vector clearly gives rise to a submodule and vise versa. Indeed, given a submodule let us consider its lowest weight (with respect to $J_{+-}$) subspace. Because $J_{-a}$ lowers the weight this subspace is annihilated by $J_{-a}$.

For the Verma module under consideration the structure is known and can be easily found by direct computation. It turns out that for $\Delta\neq \frac{d}{2}-\ell, \,\, \ell \in \NN$ (recall that we also assumed $-\Delta \notin \NN_0$) the module doesn't contain singular vectors and hence is irreducible. For $\Delta=\frac{d}{2}-\ell, \,\, \ell \in \NN$ (such values are refereed to as \textit{critical} in what follows) there is a singular vector of the form $(y^2)^\ell$ and hence a submodule of elements of the form $(y^2)^\ell f(y)$, where $f(y)$ is an arbitrary power series in $y$. This is easy to check directly using the explicit form of $J_{-a}$. The quotient module is irreducible (recall that $-\Delta \notin \NN$).

Because as a linear space the entire Verma module can be identified with all formal series in $y^a$, the quotient module is isomorphic to those series which are in the kernel of $(\dl{y^a}\dl{y_a})^\ell$. Indeed, passing to the graded dual module (using a usual inner product on homogeneous polynomials) one finds that the quotient is mapped to a submodule of elements annihilated by $(\dl{y^a}\dl{y_a})^\ell$.

\subsection{Parent formulation}

\label{sec:scalar-parent}

Still using ambient scalar~\eqref{ads-scalar} as an example let us recall the parent extension~\cite{Bekaert:2012vt} of the ambient space approach to boundary values. We closely follow~\cite{Bekaert:2012vt} to which we refer for further details. Let us introduce new variables $Y^A$ and consider an extended system:
\begin{equation}
\label{parent-ambient}
 (\dl{X^A}-\dl{Y^A})\Phi=0\,, \qquad \dl{Y}\cdot\dl{Y}\Phi=0\,, \qquad ((X+Y)\cdot\d_Y+\Delta)\Phi=0\,,
\end{equation} 
where $\Phi$ is now allowed to depend on $Y$. It is easy to check that this system is equivalent to~\eqref{ads-scalar} via elimination of auxiliary fields provided $\Phi$ depends on $Y$ formally. In other words $\Phi$ is a generating function for fields $\Phi^0,\Phi^1_A,\Phi^2_{AB},\ldots$ which are the expansion coefficients.

The first equation in~\eqref{parent-ambient} can be understood as a covariant constancy condition determined by a particular $iso(d,2)$ connection. Namely, the one where $E^A=dX^A$ and $\omega^{AB}=0$ (here we identify the connection components as the ambient frame field and Lorentz connection). Interpreted in this way the above equations can naturally be rewritten using generic coordinates $X^\ul{A}$ on the ambient space and generic local frame of the tangent bundle. The system takes the following form
\begin{equation}
\label{parent-amb}
 \nabla\Phi=0\,, \qquad \dl{Y}\cdot\dl{Y}\Phi=0\,, \qquad \left((V({X})+Y)\cdot\d_Y+\Delta\right)\Phi=0\,
\end{equation} 
which is refereed to as the parent form of the ambient system.
Here 
\begin{equation}
\nabla=\derham-E^A\dl{Y^A}- \omega^B_A \,Y^A\dl{Y^B}\,,
\end{equation} 
where $\derham=d{X}^{\ul{C}}\dl{{X}^{\ul{C}}}$ is the De Rham differential, $\omega^A_B=d{X}^\ul{C}\omega_{\ul{C} B}^A$ and $E^A=d{X}^\ul{C}E_{\ul{C}}^A$ are components
of a flat $iso(d,2)$ connection, and $V^A({X})$ are components of the section which in the suitable local frame coincide with the starting point Cartesian coordinates $X^A$. In particular, $V\cdot V=X\cdot X$. The compatibility conditions are
\begin{equation}
 \derham \omega^A_B+\omega^A_C\omega^C_B=0\,, \qquad \derham V^A+\omega^A_B V^B=E^A\,.
\end{equation} 
One can show that given a flat $iso(d,2)$-connection and a section $V^A$ satisfying the above conditions one can choose a local frame and local coordinates ${X}^{A}$ such that $V^A={X}^A, \omega^A_B=0, E^A_B=\delta^A_B$. The geometric idea behind the system~\eqref{parent-amb}  is to use the ambient space construction in the tangent space rather than in spacetime.

It is easy to consider the parent form of the ambient system~\eqref{parent-amb} as the system defined on the hyperboloid or the conformal space. For instance, by simply pulling back the ambient tangent bundle to a submanifold $X^2=-1$ one finds the system defined explicitly on $X^2=-1$. This is equivalent  to considering the original ambient system~\eqref{ads-scalar} in the vicinity of the hyperboloid $X^2=-1$. The resulting system is now determined by the same equations~\eqref{parent-amb} except that $\Phi$ is defined on $X^2=-1$,  $\omega^A_B,E^A$ and $V^A$ are components of respectively the connection and the nonvanishing section $V$
defined on the hyperboloid. Note that now $V^2=-1$. One can check that as local field theories 
defined on the hyperboloid the parent system and the starting point scalar field  $(\nabla^2-m^2)\phi=0$ are equivalent through the elimination of generalized auxiliary fields. 

Because $V^2=-1$ on the hyperboloid one can use a local frame where $V^A$ are constant.
Note that in such frame a covariant derivative takes the following form
\begin{equation}
\label{nablaJ}
 \nabla=\derham-\omega_A^B(V^A+Y^A)\dl{Y^B}
\end{equation} 
and hence can be regarded as that of a flat $o(d,2)$ connections in the associated vector
bundle, where $o(d,2)$ acts in the fiber as $J_{AB}=(V_A+Y_A)\dl{Y^B}-(V_B+Y_B)\dl{Y^A}$, i.e. $\nabla=\derham+\half\omega^{AB}J_{AB}$. In a more special frame, where $V^{d+1}=1, V^{m}=0, m=0,\ldots, d$ one recovers a framework of Section~\bref{sec:weyl-ads}. In particular, the constraints in~\eqref{parent-amb} are precisely~\eqref{ads-conf-scalar}. Having solved them one arrives at the unfolded formulation $\nabla\psi=0$ where $\psi$ takes values in the subspace~\eqref{ads-conf-scalar}, i.e. the Weyl module discussed in~\bref{sec:weyl-ads}. This unfolded formulation was arrived at in~\cite{Shaynkman:2000ts} from a rather different perspective.

Repeating the same steps for the conformal space $\manX$, identified as a submanifold of the hypercone $X^2=0$ one arrives at the parent formulation in terms of fields defined on $\manX$. A convenient choice
of the local frame is again such that $V^A=const$ and, just like above, the covariant derivative has the structure~\eqref{nablaJ}. Picking $V^+=1,V^-=V^a=0$ and explicitly solving the algebraic constraints we reproduce the framework of Section~\bref{sec:Weyl-conf} as well as the unfolded formulation of this conformal system. For $-\Delta\notin \NN$ such unfolded formulation
was proposed in~\cite{Shaynkman:2004vu} from the representation-theoretical perspective.

Furthermore, parent formulation on $\manX$ can be considered a generating procedure for the equations satisfied by the boundary values. To demonstrate this let us assume that $\Phi$ is defined on $\manX$ and pick a local coordinate system $x^a$ on $\manX$ and the local frame such that the only nonvanishing components of the flat connection $\omega$ are $\omega_+^a=dx^a$, $\omega_a^-=-dx_a$ so that the covariant derivative reads as
\begin{equation}
 \nabla=dx^a(\dl{x^a}-(Y^++1)\dl{y^a}+y_a\dl{u})\,,
\end{equation} 
where $u\equiv Y^-$. 

Now the system \eqref{parent-amb} takes the form
\begin{equation}
\label{conf-amb-sc}
 \nabla \Phi=0\,, \quad (\dl{Y^+}\dl{u}+\dl{y^a}\dl{y_a})\Phi=0\,, \quad  (\dl{Y^+}+Y\cdot \dl{Y}+\Delta) \Phi=0\,,
\end{equation} 
The first and the third equations are first-order in $y^a$ and $Y^+$ derivatives and hence have a unique solution for a given initial condition $\phi(x,u)=\Phi|_{y^a=Y^+=0}$. In terms of $\phi$ the second equation implies (for more details see~\cite{Bekaert:2012vt})
\begin{equation}
\label{scalar-ord-der}
\Box_0 \phi+\dl{u}\left(d-2\Delta-2u\dl{u}\right)\phi=0\,.
\end{equation} 
This equation does not impose any constraints on $\phi_0=\phi|_{u=0}$ for $\Delta\neq \frac{d}{2}-\ell$ with $\ell \in \NN$. However, if $\Delta=\frac{d}{2}-\ell, \ell \in \NN$ then $\phi_0$ is subject to $\Box_0^\ell\phi_0=0$. In other words, in that case \eqref{conf-amb-sc} is equivalent through the elimination of auxiliary fields to the equation $\Box_0^\ell\phi_0=0$ on two scalar fields $\phi_0$ and $\phi_\ell$ (i.e. $\phi_\ell$ is unconstrained, it is related to a subleading boundary value and is the $\ell$-th coefficient in the expansion of $\phi$ in powers of $u$). If in addition to the constraints one also takes into account the $Y$-space version $\Phi\sim \Phi+ ((V+Y)\cdot(V+Y))^{\ell}\alpha$ of the gauge equivalence from~\eqref{conf-scalar}, the subleading $\phi_l$ is eliminated and the parent form of the complete system \eqref{conf-scalar} is equivalent to just $\Box_0^\ell \phi_0=0$. 

Of course this analysis is equivalent to the standard near boundary analysis~\cite{FG,Skenderis:2002wp} (see also~\cite{Gover:2011rz}) except that in our approach there is no room for the log term and hence the leading boundary value is constrained). 
However, as a byproduct of the analysis we see that \eqref{conf-amb-sc} supplemented by
a $Y$-space version $\Phi\sim \Phi+ ((V+Y)\cdot(V+Y))^{\ell}\alpha$ of the equivalence relation from~\eqref{conf-scalar} gives a manifestly $o(d,2)$-invariant formulation of the conformal
polywave equation. Note that the intermediate equation~\eqref{scalar-ord-der} can be interpreted as an ordinary derivative formulation of the, in general higher derivative, equation $\Box_0^\ell \phi_0=0$. Formulations of this sort were developed for a rather general conformal gauge fields in~\cite{Metsaev:2007fq,Metsaev:2007rw}.

A remarkable feature of the parent approach to boundary values is that in the parent form the passage from the system describing AdS field to that describing its boundary value essentially amounts to replacing AdS compensator $V^2=-1$ by the conformal one $V^2=0$. This remains true for more general (gauge) fields. Note that the parent formulation naturally extends to gauge systems. In this case in addition to the algebraic constraints in $Y$ space, the fields are subject to the algebraic gauge equivalence relations and extra gauge fields are present in the formulation.

\section{Mixed symmetry fields and their boundary values}

\subsection{Ambient tensors and $sp(2n)\oplus o(d,2)$ Howe duality}

The standard way to describe fields on $AdS_{d+1}$ space in such a way that the isometry
algebra is realized linearly is to work with tensors of AdS algebra instead of Lorentz tensors.
Analogous considerations apply to conformal fields in $d$ dimensions because the $d$ dimensional conformal space can be identified as a projectivization of the null hypercone in $\fR^{d,2}$.

A convenient way to work with $o(d,2)$ tensors is to introduce commuting variables $P^A_I$, where $A=0,\ldots,d+1$ is an $o(d,2)$ vector index and $I=0,\ldots, n-1$ where $n\leq [\frac{d}{2}]$. The space of functions in $P^A_I$ is naturally an $o(d,2)-sp(2n)$-bimodule. $o(d,2)$ acts as
\begin{equation}
J_{AB} = P_{IA} \frac{\d}{\d P^B_{I}} - P_{IB} \frac{\d}{\d P^A_{I}}
\end{equation}
and $sp(2n)$ acts as
\begin{equation}
T_{IJ} = P^A_I P_{JA}, \qquad T_I{}^J = \frac12 \{P^A_I, \frac{\d}{\d P^A_J}\}, \qquad T^{IJ} = \frac{\d}{\d P^A_I} \frac{\d}{\d P_{JA}}.
\end{equation}
These two algebras commute in this representation. They form a dual pair $o(d,2)-sp(2n)$ in the sense of Howe~\cite{Howe1}.

In application to AdS and/or conformal fields it is useful to distinguish variables with $I=0$ and the remaining variables $P^A_i$, $i=1,\ldots,n-1$. More precisely, $P^A_0$ are to be identified with coordinates $X^A$ on the ambient space $\fR^{d+2}$. Accordingly, instead of polynomials in $P^A_I$ it is natural to consider polynomials in $P^A_i$ with coefficients in smooth functions on $\mathbb R^{d,2}$ with the origin excluded.

In what follows we use the following notation for some of the $sp(2n)$ generators
\begin{equation}
\label{sp2nnotations}
\begin{gathered}
\Box = \frac12 \d^X \cdot \d_X, \qquad\quad  \barBox = \frac12 X^2,\qquad\quad 
S^i = \d_P^i \cdot \d_X, \qquad\quad \bar S_i = P_i \cdot X,\\
S^\dag_i = P_i \cdot \d_X, \qquad\quad \bar S^\dag{}^i = X \cdot \d_P^i,\qquad\quad
T^{ij} = \d_P^i \cdot \d_P^j, \qquad\quad \bar T_{ij} = P_i \cdot P_j,\\
{N_i}^j = P_i \cdot \d_P^j, \qquad\quad N_i = {N_i}^i \quad \text{(no summation)},\\
N_X 
= X \cdot \d_X,\quad\qquad A = 0, \ldots, d+1,\qquad i = 1, \ldots, n-1,
\end{gathered}
\end{equation}
where $o(d,2)$-indices are implicit, $\d_X$ and $\d_P^i$ stand for $\dl{X^A}$ and $\dl{P_i^A}$ respectively, and $A\cdot B$ denotes the invariant contraction of $o(d,2)$ indices e.g. $A\cdot B=\eta_{CD}A^C B^D$.

\subsection{Ambient description of (partially) massless fields in anti-de Sitter space}
\label{sec:review-ads-gauge-fields}

A (partially) massless field on $AdS_{d+1}$ of spin $\{s_1,s_2,\ldots,s_{n-1}\}$ (it is assumed that spin numbers are ordered as $s_1 \ge s_2 \ge \ldots \ge s_{n-1}$ and $n-1\le [\frac{d}{2}]$)  is characterized~\cite{Metsaev:1995re,Skvortsov:2009zu} by an integer $p \le n-1$ (number of "gauge lines" in the Young diagram) and a positive integer $t \le s_p - s_{p+1}$. Here we recall the ambient formulation of~\cite{Alkalaev:2009vm,Alkalaev:2011zv} (see also~\cite{Grigoriev:2011gp,Joung:2011ww,Joung:2012rv} and earlier related works~\cite{Bonelli:2003zu,Barnich:2006pc,Hallowell:2005np,Fotopoulos:2008ka}) where the field is described by the constraints and the gauge equivalence relation which are (expressed through) the $sp(2n)$ algebra generators~\eqref{sp2nnotations} and hence the formulation is manifestly $o(d,2)$-invariant.

The field is encoded in the generating function $\Phi(X,P)$ which is an ambient space function with values in polynomials in $P^A_i$, $A=0,1,\ldots,d+1$, $i=1,\ldots,n-1$. The constraints to be imposed on the ambient space field $\Phi(X,P)$ can be grouped as follows:
\paragraph{Purely algebraic constraints:} these are generalized tracelessness, Young-symmetry and spin-weight conditions:
\begin{equation}
T^{ij} \Phi = 0, \qquad N_i{}^j \Phi = 0, \; i<j, \qquad N_i \Phi = s_i \Phi.
\end{equation}

\paragraph{(Generalized) tangent constraints:}
\begin{equation} \label{ambient_higher_power}
(\bar S^\dagger{}^p)^t \Phi = 0\,, \qquad {\bar S^\dagger}{}^{\hat \alpha}\Phi=0\,,\quad \hat \alpha=p+1,\ldots,n-1
\end{equation}
their role is to reduce tensor in $d+2$ dimensions to (a collections of) tensors in $d+1$. 

\paragraph{Radial weight constraint:}
\begin{equation}
(N_X + \Delta_\Phi) \Phi = 0,
\end{equation}
where  $\Delta_\Phi = t + p - s_p$.  Thanks to this one the field configurations in the ambient space are one to one with the configuration on the hyperboloid. More technically, in a suitable coordinate system $x^\mu,r=\sqrt{-X^2}$ such that $X^A\cdot \dl{X^A} x^\mu=0$ (i.e. $x^\mu$ can be seen as local coordinates on the hyperboloid) this constraint allows one to uniquely determine the $r$ dependence and hence to express the ambient field in terms of its value at $r=1$. Although this constraint contains derivatives in $X^A$  as a matter of fact it doesn't produce differential constraints for $\Phi$ on the hyperboloid. Indeed $X^A\cdot \dl{X^A}$
is transversal to the hyperboloid.

\paragraph{Equations of motion and partial gauges:} 
\begin{equation}
\Box \Phi = 0, \qquad S^i \Phi = 0.
\end{equation}
In contrast to the above ones these are essentially differential constraints because they do involve $X^A$ derivatives along the hyperboloid and, being rewritten in terms of tensor fields on the hyperboloid, are precisely the equations of motion together with partial gauge conditions.

\paragraph{Gauge invariance.} The gauge transformation is given by
\begin{equation}
\delta_\chi \Phi = S^\dag_\alpha \chi^\alpha, \qquad \alpha = 1, \ldots, p,
\end{equation}
where gauge parameters $\chi^\alpha$ satisfy the same constraints as $\Phi$ except those involving $N_X, N_i, N_i{}^j$ which are replaced by
\begin{equation}
(N_X + \Delta_\chi) \chi^\alpha = 0, \qquad \Delta_\chi = \Delta_\Phi - 1,
\end{equation}
\begin{equation}
N_i \chi^\alpha = s_i \chi^\alpha - \delta^\alpha_i \chi^\alpha,
\end{equation}
\begin{equation} \label{ambient_Young}
N_i{}^j \chi^\alpha = - \delta^\alpha_i \delta^j_\beta \chi^\beta, \qquad i < j.
\end{equation}

Note that gauge parameters are dependent. There is only one independent parameter as can be directly seen from \eqref{ambient_Young}: $N_1{}^i \chi^1 = -\chi^i, \; i > 1$. For later purposes it can be useful to express them through $\chi^p$. Namely
\begin{equation}
  \chi^i = - \frac{1}{s_i - s_p + 1} N_p{}^i \chi^p.
\end{equation}
So the gauge transformation takes the form (the gauge transformations for the mixed symmetry fields were originally found~\cite{Metsaev:1995re} in this form)
\begin{equation}
  \delta_\chi \Phi = \left( S^\dag_p - \frac{1}{s_{p-1} - s_p + 1} S^\dag_{p-1} N_p{}^{p-1} - \ldots - \frac{1}{s_1 - s_p + 1} S^\dag_1 N_p{}^1 \right) \chi^p.
\end{equation}

It follows from the dependence of the gauge parameters that the gauge symmetry is not irreducible for $p>1$. This means that the definition of the gauge system should also involve specification of reducibility relations (also known as gauge generators for gauge parameters) and reducibility parameters (also known as gauge for gauge or higher level gauge parameters). For instance, it is natural to regard gauge parameter $\chi^\alpha$ as trivial (pure gauge) if it can be represented in the form $\chi^\alpha=\sd_\beta \chi_{(2)}^{\beta \alpha }$ for some $\chi_{(2)}^{\alpha \beta}=-\chi_{(2)}^{\beta\alpha}$.  It is clear that continuing the same way one finds precisely $p$ levels of degeneracy such that the level-$k$ parameter is a totally antisymmetric $\chi_{(k)}^{\alpha_1\ldots \alpha_k}$. In the next section we provide a concise description of the complete structure of the gauge symmetries using the BRST first-quantized framework.

The description is somewhat simplified if $s \equiv s_1 = s_2 = \dots = s_p$ . If in addition $t=1$ and $p$ is not such that{}\footnote{Fields of this type, e.g. the rank-2 antisymmetric gauge field in $AdS_5$, are rather peculiar, see e.g.~\cite{Ferrara:1998bv}, and we refrain from considering them. Massive fields of this sort are known as self-dual and were considered in~\cite{Metsaev:2004ee}.} $d=2p$ the field is unitary~\cite{Metsaev:1995re}. In particular, $N_\alpha{}^\beta \Phi = 0$ for any $\alpha \ne \beta$.  To see this note that $N_\alpha - N_\beta, N_\alpha{}^\beta, N_\beta{}^\alpha$ generate ${sl}(2)$ subalgebra for fixed $\alpha,\beta$ satisfying $\alpha<\beta$:
\begin{equation}
\begin{split}
&[N_\alpha{}^\beta, N_\beta{}^\alpha] = N_\alpha - N_\beta,\\
&[N_\alpha - N_\beta, N_\alpha{}^\beta] = 2 N_\alpha{}^\beta,\\
&[N_\alpha - N_\beta, N_\beta{}^\alpha] = -2 N_\beta{}^\alpha.
\end{split}
\end{equation}
It follows $\Phi$ can be regarded as a highest weight vector (i.e. annihilated by $N_\alpha{}^\beta$ with $\alpha<\beta$) of vanishing weight. This in turn implies that $\Phi$ is a lowest weight vector as well. In this case the tangent constraint $\bsd{}^p\Phi=0$ also imply $\bsd{}^\alpha \Phi = 0, i = 1, \ldots, n-1$, thanks to the algebra ($[N_\alpha{}^\beta,\bsd{}^\gamma] = \delta^\gamma_\alpha \bsd{}^\beta$). And finally, the tangent constraints can be simply written as $\bsd{}^i\Phi=0$.

\subsection{BRST first-quantized formulation}

The gauge (for gauge) symmetries of the mixed-symmetry fields can be encoded in the following BRST operator
\begin{equation}
\label{Q}
Q = S^\dag_\alpha \frac{\d}{\d b_\alpha}
\end{equation}
which is defined on the space of functions $\Psi(X,P|b)$ regarded as functions in the ambient coordinates $X^A$ taking values in the polynomials in $P^A_i$ and fermionic ghost variables $b_\alpha$, $\gh{b_\alpha} = -1$.

The field $\Phi$ considered above is identified as the ghost degree zero component of $\Psi$, gauge parameters are identified with the ghost number $-1$ component, and the order-$k$ reducibility parameters are found at ghost degree $-k$. Namely, the decomposition of the generating function $\Psi$ with respect to ghost variables reads as
\begin{equation}
\Psi = \Phi(X,P) + b_\alpha \chi^\alpha(X,P)+\sum_{k=2}^p \chi_k^{\alpha_1\ldots \alpha_k}b_{\alpha_1}\ldots b_{\alpha_k}\,,
\end{equation}

The constraints for the generating function $\Psi$ are analogous to those of $\Phi$ except for $N_X, N_i, N_i{}^j$ that have to be replaced by their $Q$-invariant extensions:
\begin{equation}
\begin{gathered}
\widehat N_i{}^j = N_i{}^j + \delta^\alpha_i \delta^j_\beta b_\alpha \frac{\d}{\d b_\beta},\qquad
\widehat N_X = N_X - b_\alpha \frac{\d}{\d b_\alpha}\,,\qquad \widehat N_i=\widehat N_i{}^i\,.
\end{gathered}
\end{equation}
In terms of the components these constraints reproduce those for fields, gauge parameters, and (higher order) reducibility parameters. Because $Q$ preserves the constraints (and hence the subspace they single out) the system is consistent. The above BRST formulation has been proposed in~\cite{Alkalaev:2009vm,Alkalaev:2011zv} to which we refer for further details. General exposition of the BRST first-quantized approach can be found e.g. in~\cite{\BGST,\BGadS}.

\subsection{Boundary values and manifestly conformal formulation}

Now we restrict ourselves to the case of unitary gauge fields. According to the discussion at the end of Section~\bref{sec:review-ads-gauge-fields} in this case $s \equiv s_1 = s_2 = \dots = s_p$, $\,t=1$, and $p$ is such that $2p\neq d$.

The  entire set of constraints can be rewritten in terms of $\Psi(X,P,b)$ as
\begin{equation}
\label{unitary-const}
\begin{gathered}
T^{ij} \Psi = 0, \qquad \hat N_i{}^j \Psi = 0, \; i<j, \qquad \hat N_i \Psi = s_i \Psi,\\
\Box \Psi = 0, \qquad S^i \Psi = 0,\qquad  \bar S^\dag{}^i \Psi = 0\,,\\
(\hat N_X + \Delta_\Phi) \Psi = 0, \quad \Delta_\Phi = 1 + p - s_p,
\end{gathered}
\end{equation}
As we have seen all the information on gauge invariance is encoded in $Q$. It is straitforward to check that $Q$ is well defined on the above subspace.

According to the general discussion of Section~\bref{sec:bound-gen} the description of the boundary values is achieved by considering the above constrained system in the vicinity of the hypercone $X^2=0$ rather than the hyperboloid $X^2=-1$. The resulting system describes boundary data and in general encodes both the leading and the subleading boundary values. It turns out that the subleading can be factored out already at the level of the above ambient constrained system, just like in~\eqref{conf-scalar} in the case of the scalar field. In the case of totally-symmetric fields this was shown in~\cite{Bekaert:2013zya}. The crucial point is that such a factorization is consistent with the gauge symmetry in the sense that both the field and the gauge parameters are factorized and the gauge generator is well-defined on the quotient. As a byproduct, in this way one gets~\cite{Bekaert:2013zya} a manifestly conformal ambient description of totally-symmetric conformal gauge fields. In this case these are Fradkin--Tseytlin fields~\cite{Fradkin:1985am,Segal:2002gd} and their higher-depth generalizations~\cite{Deser:1983mm,Erdmenger:1997wy,Vasiliev:2009ck,Bekaert:2013zya,Beccaria:2015vaa}.

Now we are interested in mixed-symmetry fields. A substantial difference with the totally-symmetric ones is that the respective gauge system is essentially reducible and hence along with the fields and the gauge parameters the reducibility parameters are present. Accordingly, the factorization procedure should extend to reducibility parameters as well. As usual a powerful technique to work with general gauge systems is the above BRST formulation where fields, gauge parameters, and reducibility parameters are different components of one and the same generating function $\Psi(X,P,b)$.

Let us quotient the space \eqref{unitary-const} of $\Psi$-configurations over the subspace of configurations of the form 
\begin{equation}
\barBox^{\hat \ell} \alpha\,, \qquad  \hat\ell = \ell+b_\alpha\dl{b_\alpha},\quad \ell=\frac{d}{2} + s_p - p -1\,,
\end{equation}
(note that $\Delta_\Phi = \frac{d}{2} - \ell$ and for a unitary field $\ell\geq 1$), where $\alpha=\alpha(X,P,b)$ satisfies the same constraints as $\Psi$ except for the radial one that becomes
\begin{equation}
\label{alphaweight}
(N_X + \frac{d}{2} + \hat \ell) \alpha = 0\,.
\end{equation}
This is consistent because $\barBox^{\hat \ell} \alpha=0$ satisfies exactly the same constraints as $\Psi$.
Indeed, it is easy to check that $\Box \barBox^{\hat\ell} \alpha=0$ thanks to~\eqref{alphaweight}. Moreover, $\bar S^\dag{}^i\barBox^{\hat\ell} \alpha=T^{ij}\barBox^{\hat\ell} \alpha=N_i^j\barBox^{\hat\ell} \alpha=0$ and the remaining constraints are satisfied thanks to the constraint algebra.

It turns out that $Q$ is well defined on the quotient space:
\begin{equation}
\label{equiv}
 \Psi \sim \Psi +\barBox^{\hat\ell} \alpha\,.
\end{equation} 
To see this it is enough to check that 
\begin{equation}
Q\barBox^{\hat\ell} \alpha = \barBox^{\hat\ell} \beta(\alpha) \,,                                                           \end{equation}
for some $\beta(\alpha)$ satisfying the same constraints as $\alpha$. Direct computation show that
\begin{equation}
 \beta(\alpha)=\hat l (X\cdot P_\gamma)\dl{b_\gamma}\alpha+\barBox Q\alpha\,.
\end{equation} 
To see that $\beta(\alpha)$ indeed satisfies all the constraints one can of course perform direct check. It is more instructive, however, to first observe that $(N_X + \frac{d}{2} + \hat \ell) \beta(\alpha)=0$. Then, upon acting with $\Box$ on both sides of $Q\barBox^{\hat\ell} \alpha = \barBox^{\hat\ell} \beta(\alpha)$ one gets zero in the LHS  and $\barBox^\ell\Box \alpha(\beta)$ in the RHS (note that the term with $\commut{\Box}{\barBox^\ell}$ vanishes
thanks to $(N_X + \frac{d}{2} + \hat \ell) \beta(\alpha)=0$). Taking into account that the kernel of $\barBox^{\ell}$ is trivial one concludes that $\Box\beta(\alpha)=0$. Analogously one finds $\bsd{}^i\beta(\alpha)=0$. The rest follows from the constraint algebra.

In particular, for $b_\alpha$-independent component $\Phi$ the factorization is performed with respect to configurations of the form $\barBox^\ell \alpha_0$.  For linear in $b_\alpha$ component of $\Psi$ one finds factorization of gauge parameters over the subspace of elements of the form $\barBox^{\ell+1} \alpha_1^\gamma$. Namely $\chi^\gamma \sim \chi^\gamma + \barBox^{\ell+1}\alpha_1^\gamma$. One can show that this is the most general factorization consistent with the gauge transformations.

We have thus constructed the space \eqref{unitary-const}, \eqref{equiv} of configurations for fields and (higher level) gauge parameters, which is equipped with the BRST operator~\eqref{Q}. This gives a manifestly conformal formulation of the equations of motion, gauge symmetries and (higher order) reducibility relations. Indeed, both the space  and the BRST operator are defined in a manifestly $o(d,2)$-invariant way because only $o(d,2)$-invariant operators enter the defining relations for the space \eqref{unitary-const}, \eqref{equiv} and $Q$ \eqref{Q}. In this context in addition to the manifestly conformal formulation~\cite{Bekaert:2009fg} of higher-spin singleton fields let us also mention the formulations of~\cite{Arvidsson:2006fq,Marnelius:2008er}.

\subsection{Parent formulation and elimination of auxiliary fields}
\label{sec:parent-mixed}

Although the conformal symmetry is manifest in this formulation, the system is not explicitly given in terms of fields defined on the conformal space, and, in addition, the gauge and reducibility parameters are subject to some differential constraints. Both these problems can be cured by passing to the parent formulation.

In contrast to a system without gauge symmetries (e.g. the one considered in Section~\bref{sec:scalar-parent}), the passage to the parent formulation for a gauge system doesn't only boil down to adding $Y$-variables and the covariant constancy constraint. However, the required generalization is straightforward if one keeps using the BRST framework.

More precisely, the generating function for fields and (gauge for) gauge parameters is $\Psi(x,\theta|Y,P,b)$ where $x^\mu$ are coordinates on $\manX$ and $\theta^\mu$ are basis differentials $dx^\mu$ seen as Grassmann odd ghost variables with $\gh{\theta^\mu}=1$. $\Psi$ is subject to the constraints \eqref{unitary-const}-\eqref{equiv} and the equivalence relation \eqref{equiv} where the following replacements has been made $\dl{X^A}\to \dl{Y^A}$ and  $X^A \to V^A+Y^A$. The parent BRST operator reads as
\begin{equation}
\begin{gathered}
\Omega=\nabla+\bar Q\,,  \qquad \qquad\nabla=\derham+\half\omega^{AB}J_{AB}\,, \\
\bar Q=Q\Big|_{{\frac{\d}{\d X}\to \frac{\d}{\d Y}}, X \to V+Y}=(P_\alpha\cdot\dl{Y})\dl{b_\alpha}
\end{gathered}
\end{equation} 
where $J_{AB}$ denote the  $d(d,2)$-generators in the twisted representation:
\begin{equation}
 J_{AB}=(V_A+Y_A)
 \dl{Y^B} -(V_B+Y_B)
 \dl{Y^A} +P_{iA}
 \dl{P^{B}_i} -P_{iB}
 \dl{P^{A}_i} \,.
\end{equation} 
Note that the replacements made do not affect the commutation relations satisfied by $sp(2n)$ and $o(d,2)$ generators. 

The standard prescription of the BRST formulation is that physical fields are contained in the ghost degree zero component $\Phi$ of $\Psi$ while components of negative degree are interpreted as gauge (for gauge) parameters. In particular, physical fields are
\begin{equation}
 \Phi=\phi(x,Y,P)+\theta^\mu \phi_{\mu}^\alpha(x,Y,P)b_\alpha+\theta^\mu \theta^\nu \phi_{\mu \nu }^{\alpha\beta}(x,Y,P)b_\alpha b_\beta+\ldots
\end{equation} 
so that differential forms of degree $0,1,\ldots,p$ are present. Notations for gauge (for gauge) parameters are introduced according to 
\begin{equation}
 \Psi=\Phi+\sum_{m=1}^p \Psi^{(m)}+\ldots
  \,, \qquad \gh{\Psi^{(m)}}=-m\,,
\end{equation} 
where dots denote components of positive ghost degree which are associated to antifields.

The equations of motion and the (higher order) gauge transformations read as:
\begin{equation}
\label{parent-mixed}
 (\nabla+\bar Q) \Phi =0\,, \quad \delta \Phi = (\nabla+\bar Q)\Psi^{(1)}\,, \qquad \delta \Psi^{(m)}=(\nabla+\bar Q)\Psi^{(m+1)}\,.
\end{equation} 
Together with~\eqref{unitary-const-twist} these define the parent formulation of the leading boundary values. This system is explicitly defined in terms of fields on $\manX$, differential constraints on (higher level) gauge parameters are not present, and $o(d,2)$-invariance is realized in a manifest way so that the system provides gauge covariant, manifestly local
and $o(d,2)$-invariant formulation of the leading boundary values. However, for practical purposes such as deriving component formulations it can be useful to restrict the formulation
by partially fixing the gauge (for gauge) invariance.

A convenient partial gauge is $\dl{\theta_\mu}\Phi=0$ i.e. all nonzero forms $\phi_{\mu\ldots}^{\alpha\ldots}(x,Y,P)$ are put to zero. In order to preserve the gauge condition the gauge  parameters have to satisfy $(\nabla+\bar Q)\Psi^{(1)}=0$. Just like for the fields themselves one can use the gauge for gauge transformations to achieve $\dl{\theta^\mu}\Psi^{(1)}=0$, i.e. put to zero all parameters which are forms of nonvanishing degree. Requiring the second order parameters $\Psi^{(2)}$ to preserve $\dl{\theta^\mu}\Psi^{(1)}=0$ one arrives at $(\nabla+\bar Q)\Psi^{(2)}=0$ and again one can use the next level gauge transformations to achieve $\dl{\theta^\mu}\Psi^{(2)}=0$. Continuing the same way one arrives at the partial gauge condition $\dl{\theta^\mu}\Psi=0$ for all fields and (higher level) gauge parameters contained in~$\Psi$. The residual (higher level) gauge transformations read as
\begin{equation}
\label{par-gauge}
 \delta \Phi=\bar Q \Psi^{(1)}\,, \qquad \delta \Psi^{(k)}=\bar Q \Psi^{(k+1)}, \quad k=1,\ldots p-1\,. 
\end{equation}

Let us list all the conditions imposed on fields and residual (higher level) gauge parameters:
\begin{equation}
\label{unitary-const-twist}
\begin{gathered}
\nabla \Psi=0\,,\qquad \dl{\theta^\mu}\Psi=0\\
T^{ij} \Psi = 0, \qquad  \hat N_i{}^j \Psi = 0, \; i<j, \qquad \hat N_i \Psi = s_i \Psi,\\
\d_Y\cdot \d_Y \Psi = 0, \qquad \d_Y\cdot \d^i_P \Psi = 0,\qquad  (Y+V)\cdot\d_P^i \Psi = 0\,,\\
((Y+V)\cdot \d_Y + \frac{d}{2}-\hat l) \Psi = 0\,, \qquad \hat\ell = b_\alpha\dl{b_\alpha}+\frac{d}{2} + s_p - p -1\,,\\
\Psi \sim \Psi + ((Y+V)\cdot(Y+V))^{\hat l }\alpha 
\end{gathered}
\end{equation}
where $\alpha=\alpha(x,Y,P,b)$ are subject to the same constraints but with $\hat l$ replaced by $-\hat l$. These are the $Y$-space versions of the constraints~\eqref{unitary-const} and the equivalence relation~\eqref{equiv} respectively. In the case of totally-symmetric fields this partially gauge fixed system was introduced in~\cite{Bekaert:2013zya}.

In contrast to the parent system~\eqref{parent-mixed}, the partially gauge-fixed system~\eqref{unitary-const-twist} again contains differential constraints on the gauge parameters (through $\nabla \Psi=0$). So, at first glance we are back to the same problem as in the original ambient system~\eqref{unitary-const}, \eqref{equiv}. This is true, but only formally so. In fact, as we are going to see shortly, by eliminating the auxiliary fields from the system \eqref{unitary-const-twist}, one ends up with genuine gauge invariant equations supplemented with partial gauge conditions. In other words, this system can be considered as a sort of generating procedure which allows one to find component expressions for the gauge invariant equations and associated (higher level) gauge transformations.

To illustrate this point it is instructive to start with the simplest nontrivial example of a gauge system: the spin $1$ gauge field. For $\Phi=A^B(X,Y)P_B$ equations \eqref{unitary-const-twist} take the form
\begin{equation}
\label{spin-1-parent}
 \begin{gathered}
\d_Y\cdot \d_Y \Phi=0, \qquad \d_Y\cdot \d_P \Phi=0, \\
\nabla \Phi=0,  \qquad ((Y+V)\cdot \d_Y + 1)\Phi=0, \qquad (Y+V)\cdot \d_P \Phi=0\,,                                                                                         \end{gathered}
\end{equation} 
where for the moment we omit the equivalence relation.

Now we take the compensator field, $o(d,2)$ connection, and local coordinates $x^a$ as in Section~\bref{sec:scalar-parent} i.e. $V^+=1,V^-=V^a=0$ and the only nonvanishing components of the connections are $\omega_+^a=dx^a$, $\omega_a^-=-dx_a$. Consider the equations of the second line in\eqref{spin-1-parent}: the 1st equation can be used to eliminate $y^a$, the 2nd to eliminate $Y^+$ and the 3rd to eliminate $P^+$. Upon the elimination the equations of the first line take the form
\begin{equation}
\label{s1-ord-der}
\begin{aligned}
\tilde\Box \phi+\dl{u}\left(d-2-2u\dl{u}\right)\phi&=0\,,\\
\left(\dl{p}\cdot\dl{x}\right)\phi +\dl{w}
\left(d-1-2u\dl{u}-w\dl{w}\right)
\phi&=0\,,
\end{aligned}
\end{equation}
where $u\equiv Y^-$, $w\equiv P^-$, and $\tilde\Box=\Box_0+2(p\cdot\dl{x})\dl{w}$.

Introducing notations $\phi_0=\phi|_{u=0}$ and $\phi_{00}=\phi_0|_{w=0}$ the second equation gives $\phi_0=\phi_{00}-\frac{1}{d-2}w(\d_p\cdot \d_x)\phi_{00}$.
The first equation then gives $(\tilde \Box)^{\ell}\phi_0=0$, where $\ell=\frac{d}{2}-1$ (here and below we assume $d$ even). There are two components in this equation: $w$-independent and linear in $w$. They read respectively as:
\begin{equation}
 \Box_0^{\ell-1}\left(\Box_0-(p\cdot \d_x)(\d_p\cdot \d_x)\right)\phi_{00}=0\,, \qquad \Box_0^{\ell}(\d_p\cdot \d_x)\phi_{00}=0\,.
\end{equation} 
Recalling that $\phi_{00}(x,p)=A^a(x)p_a$ one observes that the first equation is precisely the conformal spin-1 equation while the second is the respective conformal gauge condition. In fact the system \eqref{s1-ord-der} can be slightly modified to describe just the first equation:
\begin{equation}
\label{higher-order-spin1}
\left((\tilde\Box)^\ell \phi_0\right)\Big|_{w=0}=0\,,\qquad
\left(\d_{p}\cdot\d_{x}\right)\phi_0 +\dl{w}
\left(d-1-w\dl{w}\right)
\phi_0=0\,.
\end{equation} 
The second equation serves as a constraint which uniquely expresses $\phi_0(x,p,w)$ through $\phi_{00}(x,p)$. It turns out that more general conformal gauge invariant equations can be written in a similar way.

Just like in the case of the scalar field (see Section~\bref{sec:scalar}), equations~\eqref{s1-ord-der} contain the ordinary derivative formulation of the conformal spin 1 equations analogous to that of~\cite{Metsaev:2007rw,Metsaev:2007fq}. Strictly speaking~\eqref{s1-ord-der} also encode
conformal gauge condition but it can be removed by restricting the RHS of the first equation to $w=0$. Note that one may also wish to restrict $\phi$ not to depend on $u^k$ with $k\geq \ell$ in order to eliminate the subleading solution. At the level of the system~\eqref{spin-1-parent} this corresponds to taking into account the equivalence relation $\Phi\sim\Phi+((V+Y)\cdot(V+Y))^\ell\alpha$.

Analyzing the gauge transformations in the case of totally-symmetric fields one can prove~\cite{Bekaert:2012vt} that that the equation sitting at $w$-independent component (i.e. $(\tilde\Box^\ell\phi_0)_{w=0}=0$) is gauge invariant with the differentially unconstrained gauge parameter.  This is the general feature which can be shown in a rather general setting as follows: the equations arising at degree $k$ in $w$ are of order $2\ell+k$ in $x$-derivatives while the analogous equation for the gauge parameter are of order $2\ell+2k+2$. The gauge variation of the equations for $k=0$ should vanish on the equations for gauge parameter but the gauge variation is of order $2\ell+1$  while the equation for the gauge parameter is of order $2\ell+2$ so that the equation should be gauge invariant with the differentially unconstrained parameter. Analogous arguments apply to the (higher level) gauge transformations.

In fact the conformal invariance of the equation at $(\tilde\Box^\ell\phi_0)_{w=0}=0$ can also be proved on general grounds. Indeed, the system of all the equations encoded in $(\tilde\Box^\ell\phi_0)=0$ is conformal by construction. It follows the conformal transformation of $(\tilde\Box^\ell\phi_0)|_{w=0}=0$ must be proportional to a combination of the equations contained in $(\tilde\Box^\ell\phi_0)=0$. However, conformal variation is of order $2\ell$ (it is enough to consider special conformal transformations which are represented on $\phi_{00}$ by operators that have the structure $K_a=1\tensor D+k_a$ where $D$ is a scalar differential operator of order $1$ while $k_a$ act on spin components but do not contain $x$-derivatives) and hence can not be compensated by other equations in  $(\tilde\Box^\ell\phi_0)=0$ because they are of order higher than $2\ell$.

As we are going to see in the next section all the conditions for the above two arguments are fulfilled in the case of leading boundary values for generic unitary massless fields.

To conclude the discussion of the general formalism, recall that in the case of scalar an important object is the Weyl module which in the parent language is a space of solutions of the algebraic constraints imposed on $\Psi$. In the case of gauge systems a proper counterpart is the space $H^0(Q)$ ($Q$-cohomology at ghost degree $0$) evaluated in the space of $\Psi(P,Y,b)$ satisfying~\eqref{unitary-const}. Indeed, $H^0(Q)$ consists of gauge-inequivalent solutions in the space of formal power series around a fixed spacetime point. However, in general $H^i(Q)$ may also be nonvanishing for $i<0$. These cohomology groups are typically finite-dimensional and are known as (higher level) global reducibility parameters (see~\cite{Barnich:2015tma} and references therein for further details).

In the case of unitary massless fields $H^i(Q)$ for $i<0$ was computed in~\cite{Alkalaev:2009vm}. The computation was based on the following observation: for $i<0$ the cocycle condition $Q\xi=0$ necessarily implies equations of the form $(P_i \cdot \dl{Y})\xi=0$ for some $i$ so that $\xi$ may involves only finite orders of $Y$-variables and hence the cohomology computation can be performed in the space of polynomials in $Y^A$. Because in this space it is legitimate to redefine $Y\to Y+V$ the result doesn't depend on $V^A$. In particular, $H^i(Q),\,\, i\leq 0$ are the same for the system on AdS and the system describing its leading boundary values. In the unfolded formalism this leads to the match between the bulk and the boundary gauge fields (for the unfolded approach to boundary values see~\cite{Vasiliev:2012vf}).

\subsection{Boundary values in terms of components}

Now we derive concise formulations of the equations of motion and gauge symmetries
for generic unitary massless mixed-symmetry field, generalizing the spin-$1$ equations~\eqref{s1-ord-der} and \eqref{higher-order-spin1}. To this end let us rewrite \eqref{unitary-const-twist} for the physical (i.e. $b,\theta$-independent) component  field $\Phi$ (omitting for the moment the equivalence relation):
\begin{gather}
\nabla \Phi=0\,, \qquad ((Y+V)\cdot \d_X + \Delta) \Phi = 0\,, 
\qquad (Y+V)\cdot\d_P^i \Phi = 0\,,\label{eq11}\\
\d_Y\cdot \d_Y \Phi = 0, \qquad \d_Y\cdot \d^i_P \Phi = 0\,,\label{eq12}\\
T^{ij} \Phi = 0, \qquad  N_i{}^j \Phi = 0, \; i<j, \qquad N_i \Phi = s_i \Phi\,.\label{eq13}
\end{gather}

As before we use Cartesian coordinates $x^a$ on $\manX$ and a local frame such that
\begin{equation}
 V^+=1,~~ V^-=V^a=0\,, \qquad \omega_+^a=-\omega_a^-=dx^a\,, ~~~ \omega^+_-=\omega_-^a=\omega^+_b=\omega^a_b=0\,.
\end{equation} 
Introducing notations $ u= Y^-, w_i=P_i^- , p_i^a = P_i^a$ the covariant derivative takes the form
\begin{equation} \label{nabla}
  \nabla_a = \hat\d_a - (Y^++1) \frac{\d}{\d Y^a} + Y_a \frac{\d}{\d Y^-}- \sum_i P_i^+ \frac{\d}{\d P_i^a} \,,
\end{equation}
where $\hat \d^a = \d^a + \sum\limits_i p_i^a \frac{\d}{\d w_i}$.

Equations~\eqref{eq11} are first order in $y^a,Y^+,P^+_i$ and have a unique solution for a given boundary data $\phi(x|p,u,w_i)$. In terms of $\phi(x|p,u,w_i)$ equations \eqref{eq12}-\eqref{eq13} take the form (see Appendix~\bref{sec:components-details} for more details)
\begin{gather} 
\label{unitary_varphi_box}
	\tildeBox \phi + \frac{\d}{\d u} \left[ d - 2 \left( \Delta + u \frac{\d}{\d u} \right) \right] \phi = 0,
	\\
 \label{unitary_varphi_div}
	(\d_{p_i} \cdot \d) \phi + \frac{\d}{\d w_i} \left( d + n_i - \Delta - 1 - 2 u \frac{\d}{\d u} \right) \phi + \sum\limits_{j \ne i} \frac{\d}{\d w_j} (p_j \cdot \d_{p_i}) \phi = 0,
\\
\label{unitary-spin}
\left( n_i + n_{w_i} - s_i \right) \phi = 0,
\\
	(p_i \cdot \d_{p_j}) \phi + w_i \frac{\d}{\d w_j} \phi = 0, \quad i < j,
	\label{unitary_varphi_Young}
	\\
\label{unitary_varphi_trace}
	(\d_{p_i} \cdot \d_{p_j}) \phi - 2 u \frac{\d}{\d w_i} \frac{\d}{\d w_j} \phi = 0.
\end{gather}
where $n_{w_i}=w_i\dl{w_i}$ and $\tildeBox = \hat \d^a \hat \d_a$.

Let $\phi_0=\phi|_{u=0}$. At $u=0$ equations~\eqref{unitary_varphi_div}-\eqref{unitary_varphi_trace} uniquely determine $\phi_0$ for a given initial data $\phi_{00}(x,p_i)=\phi_{0}|_{w_i=0}$ satisfying
\begin{equation}
\label{lorentz-irr}
 (n_i-s_i)\phi_{00}=0\,, \qquad (\d_{p_i} \cdot \d_{p_j}) \phi_{00}=0\,,  \qquad (p_i \cdot \d_{p_j})\phi_{00}=0\quad i<j\,.
\end{equation} 
Note that these are precisely the conditions that $\phi_{00}$ belongs to the irreducible
module with weights $s_1,\ldots, s_{n-1}$ of the Lorentz $o(d-1,1)$ subalgebra of $o(d,2)$.
Furthermore, equation~\eqref{unitary_varphi_box} determines $u$-dependence of $\phi$ and because $\ell\in \NN$, it imposes on $\phi_{0}$ the following equation: $(\tilde\Box)^\ell \phi_0=0$. Equation~\eqref{unitary_varphi_box} doesn't determine coefficient $\phi_{\ell}(x,p_i,w_i)$ of $u^\ell$ in terms of $\phi_{0}$. This is precisely the subleading solution which is gauged away by the equivalence relation $\Phi\sim \Phi+((V+Y)\cdot(V+Y))^\ell \alpha$.

Putting everything together, the leading boundary value is a Lorentz-irreducible (i.e. satisfying~\eqref{lorentz-irr}) field $\phi_{00}$ subject to the gauge transformation determined by~\eqref{par-gauge}. The gauge invariant conformal equations satisfied by $\phi_{00}$ can be written as
\begin{equation}
\label{mixed-comp}
\begin{gathered}
 ({\tilde\Box}^\ell\phi_0)|_{w_i=0}=0\,,\qquad  \phi_0|_{w_i=0}=\phi_{00}\,,\\
  (\d_{p_i} \cdot \d) \phi_0 + \frac{\d}{\d w_i} \left( d + s_i - \Delta - i - \sum\limits_{j \le i} n_{w_j} \right) \phi_0 + \sum\limits_{j>i} (p_j \cdot \d_{p_i}) \frac{\d}{\d w_j} \phi_0 = 0\,,
\end{gathered}
\end{equation} 
where the equations in the second line are interpreted as the constraints determining the $w_i$-dependence in a unique way. In the Appendix~\bref{sec:components-details} it is shown that these have a unique solution which can be obtained by solving the component equations in a certain order. Moreover, a solution to~\eqref{mixed-comp} where $w_i$ is not put to zero in the first equation also solves the original system~\eqref{unitary_varphi_box}-\eqref{unitary_varphi_trace} and vice versa. The arguments given in Section~\bref{sec:parent-mixed} show that the $({\tilde\Box}^\ell\phi_0)|_{w_i=0}=0$ is conformal invariant and gauge invariant with a differentially-unconstrained parameter. Thus we conclude that \eqref{mixed-comp} is the formulation of the conformal equation for the leading boundary value in terms of the minimal field content, i.e. Lorentz irreducible tensor $\phi_{00}$. Note that if one keeps $w_i$-variables the system \eqref{mixed-comp} gives a concise nonminimal formulation that can also be made ordinary-derivative by keeping some more auxiliary fields.

Because of the Equations~\eqref{unitary-const-twist} imposed on $\Psi$ the (higher level) gauge parameters contained in $\Psi$ satisfy the equations analogous to~\eqref{eq11}-\eqref{eq13}. Applying exactly the same arguments to these equations results in the analog of the system~\eqref{mixed-comp} for the (higher level) gauge parameters that in turn allows to express  gauge (for gauge) transformation in terms $w_i,u$-independent parameters.
In particular, gauge transformation of $\phi_{00}$ is given by 
\begin{equation}
	\delta\phi_{00} = \left( \sum_\alpha (p_\alpha \cdot \hat \d) \lambda^\alpha \right)\bigg|_{w_i = 0},
\end{equation}
where $\lambda^\alpha$ is a solution to the analogue of the last equation in \eqref{mixed-comp} with the  initial condition$\lambda^\alpha|_{w_i=0}=\lambda^\alpha_{00}$ such that $\lambda^\alpha_{00}b_\alpha$ satisfy the ghost-extended version (i.e. with $\hat N_i^j$) of the irreducibility conditions \eqref{lorentz-irr}. The analog of the last equation in \eqref{mixed-comp} is obtained by replacing $\phi_0 \mapsto \lambda^\alpha b_\alpha$, $\Delta \mapsto \Delta-1$, $s_\alpha \mapsto s_\alpha-1$. The procedure clearly extends to higher level gauge parameters and gauge for gauge transformations.

Let us finally discuss equations imposed on the subleading boundary value. 
In addition to the leading boundary value $\phi_{00}$ subject to the conformal 
equations~\eqref{mixed-comp} the 
system~\eqref{unitary_varphi_box}-\eqref{unitary_varphi_trace} also describes 
the subleading boundary value entering $\phi$  through $u^\ell \phi_\ell$. While 
the traces of $\phi_\ell$ are determined in terms of $\phi_0$ by means of 
\eqref{unitary_varphi_trace} its traceless component  $\psi_{0}$ is not, and the 
equations \eqref{unitary_varphi_div}-\eqref{unitary_varphi_Young}
impose certain equations on  $\psi_{00}= \psi_0|_{w_i=0}$. The detailed analysis 
of the equations in the general case remains beyond the scope of this work. In 
the case where $s_1>1$ and $s_2=\ldots=s_{n-1}=1$ it is straightforward to find 
that these can be written as
\begin{equation}
  \Pi \left(\frac{\d}{\d p_p^a} \d^a \psi_{00}\right) = 0\,,
\end{equation}
where $\Pi$ denotes the projection to the irreducible component described by the 
Young tableau $(s_1,\ldots,s_p-1,s_{p-1},\ldots,s_{n-1})$ (i.e. 
$(s_1-1,1,\ldots,1)$ in our case).
The subleading $\psi_{00}$ is interpreted as a current subject to the above 
conservation condition, which is in agreement with~\cite{Alkalaev:2012ic}, where 
the conserved currents associated to mixed-symmetry fields were studied.
Note that the equations on $\psi_{00}$ are by construction conformal invariant 
and hence can be found using the approach of~\cite{Shaynkman:2004vu}, if in addition one takes into account the 
tensor structure, derivative order, and the conformal weight of $\psi_{00}$.

\subsection{Example: ``hook''-type field}

As a concrete example let us consider the simplest nontrivial case: $d=4,s_1=2,s_2=1,p=1$, i.e. the mixed symmetry field described by a ``hook'' Young diagram with the antisymmetric gauge parameter:
$$
\Yvcentermath1\tiny\young(~~,~) \sim \tiny\young(~~,~) + \delta\,\tiny\young(~,~)
$$

The equations in the second line of~\eqref{mixed-comp} express $\phi_0$ in terms of $\phi_{00}$. The first equation then takes the form
\begin{multline}
  \Box^2 \phi_{abc}
    - \Box \d^e
      \left(
        \d_a \phi_{ebc}
      + \d_b \phi_{eac}
      \right)
      + \frac12 \Box \d^e
        \left(
          \d_a \phi_{bce}
          + \d_b \phi_{ace}
        \right)\\
    - 2 \Box \d^e \d_c \phi_{abe}
    + \frac12
      \left(
        \eta_{ab} \Box
        + 2 \d_a \d_b
      \right)
      \d^e \d^f
      \phi_{efc}\\
    - \frac14 \d^e \d^f
      \left[
        \left( \eta_{ac} \Box + 2 \d_a \d_c \right) \phi_{efb}
        + \left( \eta_{bc} \Box + 2 \d_b \d_c \right) \phi_{efa}
      \right]
    = 0\,.
\end{multline}
The gauge transformation in terms of independent gauge parameter $\lambda_{ab}=-\lambda_{ba}$ reads as
\begin{equation}
  \delta \phi_{abc} =
      \d_a \lambda_{bc}
    + \d_b \lambda_{ac}
    - \frac13 \d^e
      \left(
	2 \eta_{ab} \lambda_{ec}
	- \eta_{ac} \lambda_{eb}
	- \eta_{bc} \lambda_{ea}
      \right).
\end{equation}
This system is variational and the corresponding Lagrangian was originally proposed
in~\cite{Vasiliev:2009ck} on different grounds. It was also derived~\cite{Alkalaev:2012ic} from the Lagrangian~\cite{Brink:2000ag} of the respective bulk field. According to~\cite{Vasiliev:2009ck} the general conformal equations including those encoded in~\eqref{mixed-comp} are also Lagrangian and we expect that the Lagrangian can be written in a concise form suggested by~\eqref{mixed-comp}. 

\section{Conclusions} 
In this work we have generalized the approach of~\cite{Bekaert:2012vt,Bekaert:2013zya},
originally developed for totally-symmetric fields, to the case of unitary massless fields of mixed-symmetry type. As a starting point we employed the ambient space formulation of generic AdS fields proposed in~\cite{Alkalaev:2009vm,Alkalaev:2011zv}. The generalization is not entirely straightforward because mixed-symmetry fields are in general reducible gauge theories and one needs to describe boundary values for fields, gauge parameters and also reducibility parameters in a way compatible with the gauge/reducibility generators. Technically, the required generalization is elegantly achieved through the use of the BRST framework.

As a continuation of the present work it would be natural to generalize the approach to non-unitary mixed-symmetry fields, including the partially-massless ones. In this way one may expect to find a one to one match between the gauge fields on AdS and the conformal gauge fields on the boundary. The ambient formalism employed in the present work suggests that the AdS field and the associated conformal field(s) are just different faces of one and the same system that is naturally defined on the ambient space. The conformal (conserved) mixed-symmetry currents are also expected to fit into this picture.  Of the utmost importance is, of course, possible applications to interacting theories such as expected  relations between interactions of AdS higher-spin fields and those of the conformal ones. A well-known example is the relation between Einstein gravity in the bulk and the conformal gravity on the boundary.

\section*{Acknowledgments}
We are grateful to K.~Alkalaev, X.~Bekaert, O.~Shaynkman and R.~Metsaev for useful discussions.
The work of AC was supported by the RFBR grant 14-01-00489. 
The work of MG was supported by the Russian Science Foundation grant 14-42-00047.

\appendix

\section{Derivation of the component form}
\label{sec:components-details}

Equations~\eqref{eq11} can be solved order by order in $y^a,Y^+,P^+$ for any initial data
$\phi(x|u,w_i,p_i)$. Explicitly we only need few first orders: 
\begin{multline} \label{Phi_expansion}
  \Phi(x|Y, P) =
    \left[
      1
      - Y^+ \left(\Delta + u\frac{\d}{\d u}\right)
      - \sum\limits_i P_i^+ u \frac{\d}{\d w_i}
      + \right.\\
      + y^a \hat\d_a
      + \left. \frac12 y^a y^b \left( \hat\d_a \hat\d_b + \eta_{ab} \frac{\d}{\d u} \right)
      + \ldots
    \right]
    \phi(x|u,w,p)
\end{multline}
which are enough to derive \eqref{unitary_varphi_box}-\eqref{unitary_varphi_trace}.

Taking into account equations~\eqref{unitary-spin}-\eqref{unitary_varphi_trace} equation
\eqref{unitary_varphi_div} can be rewritten as 
\begin{equation} \label{unitary_varphi_div'}
	(\d_{p_i} \cdot \d) \phi + \frac{\d}{\d w_i} \left( d + s_i - \Delta - i - \sum\limits_{j \le i} n_{w_j} - 2 u \frac{\d}{\d u} \right) \phi + \sum\limits_{j>i}  (p_j \cdot \d_{p_i}) \frac{\d}{\d w_j} \phi = 0\,,
\end{equation}
which at $u=0$ gives the last equation of \eqref{mixed-comp}.

In order to show that the last equation of \eqref{mixed-comp} can be solved for any
initial data $\phi_{00}$ satisfying~\eqref{lorentz-irr} we first observe that 
the operator $d + s_i - \Delta - i - \sum\limits_{j \le i} n_{w_j}$ doesn't have zero 
eigenvalues on the subspace determined by~\eqref{unitary-spin}-\eqref{unitary_varphi_Young}.
Indeed, it follows from \eqref{unitary_varphi_Young} that elements with maximum $n_{w_i}$-eigenvalue $s_i$ and nonzero eigenvalue of $n_{w_j}$ with $i<j$ are vanishing. Furthermore, if an eigenspace with $n_{w_i}$-eigenvalue $m_i$ and $n_{w_j}$-eigenvalue $m_j$ vanishes so does the eigenspace with the respective eigenvalues  $m_i - 1$ and $m_j + 1$. This in turn implies that an eigenvalue of $\sum\limits_i n_{w_i}$ can't exceed $s_1$.

As for the possible eigenvalues of $d + s_i - \Delta - i - \sum\limits_{j \le i} n_{w_i}$
one finds that the lowest possible one is $d + s_i - (1+p-s_p) - i - s_1 = d-p-i + s_i-1$.
Because $n-1\leq [\frac{d}{2}]$, $p \leq n-1$ and $i\leq n-1$ one finds that the minimal value
is $0$. This happens when $p=n-1=d/2$ but this doesn't correspond to a unitary field
(indeed, $p=\frac{d}{2}$ in this case).

Because the coefficient in the last equation of \eqref{mixed-comp} doesn't vanish one can try to find a solution order by order in $w_i$. First one solves the equation with $i=n-1$ in the space of $w_j$-independent elements with $j<i$. Then one uses the solution as the initial data for the equation with $i=n-2$ and solves it to first order in $w_{n-2}$ and then again uses the equation with $i=n-1$ to obtain linear in $w_{n-2}$ term in the solution of the equation with $i=n-1$ and so on. In other words, we solve the last equations in \eqref{mixed-comp} order by order in the following $\mathbb N_0$-grading of weighted powers of $w_i$: $\deg{w_{i-1}} = s_i \deg{w_i} + 1$, $\deg{w_{n-1}} = 1$.

Given a solution $\phi_0$ to the last equation of \eqref{mixed-comp} (or equivalently
equation \eqref{unitary_varphi_div} or \eqref{unitary_varphi_div'} at $u=0$) for a given initial
data $\phi_{00}$ satisfying~\eqref{lorentz-irr} the equation \eqref{unitary_varphi_box}
can be used to uniquely reconstruct the dependence on $u$ up to order $\ell-1$. Let us show that
the resulting $u$-dependent solution still satisfies \eqref{unitary_varphi_div}. To this end
let $\phi=\sum\limits_{k=0}^{\ell-1} \frac{1}{k!} u^k \phi_k$ be the solution.
Suppose $u^k\phi_k$ satisfies \eqref{unitary_varphi_div'}, i.e.
\begin{equation*}
	\left\{
	  (\d_{p_i} \cdot \d) + \frac{\d}{\d w_i} \left( d + s_i - \Delta - i - \sum\limits_{j \le i} n_{w_j} - 2 k \right) + \sum\limits_{j>i}(p_j \cdot \d_{p_i}) \frac{\d}{\d w_j}
	\right\}
	\phi_k
	= 0\,.
\end{equation*}
Then using $\phi_{k+1} = C \tildeBox \phi_k$ and
\begin{equation*}
	\left[
  	(\d_{p_i} \cdot \d) + \frac{\d}{\d w_i} \left( d + s_i - \Delta - i - \sum\limits_{j \le i} n_{w_j} \right) + \sum\limits_{j>i} (p_j \cdot \d_{p_i}) \frac{\d}{\d w_j},
  	\tildeBox
	\right]
	=
	2 \frac{\d}{\d w_i} \tildeBox
\end{equation*}
we see that $u^{k+1}\phi_{k+1}$ also satisfies \eqref{unitary_varphi_div'}. Analogously one proves that $\phi_0$ satisfies \eqref{unitary-spin}-\eqref{unitary_varphi_trace} provided $\phi_{00}$ satisfies~\eqref{lorentz-irr}.

\footnotesize
{\small

\addtolength{\baselineskip}{-3.9pt}
\addtolength{\parskip}{-4pt}
% \bibliography{/home/maxim/Documents/HSmaster}
% \end{document}

\providecommand{\href}[2]{#2}\begingroup\raggedright\endgroup
}
\end{document}